\documentclass[aps,prl,superscriptaddress,
               twocolumn,balancelastpage,
               longbibliography,
               ]{revtex4-2}

\usepackage[colorlinks,bookmarks=false,citecolor=blue,linkcolor=blue,urlcolor=blue]{hyperref}
\usepackage[all]{hypcap}   % let hyperlinks correctly point to figures rather than their captions;

\usepackage{amsmath,amssymb}
\usepackage{graphicx}
\graphicspath{{Figures/}{/}}

\usepackage{verbatim}
\usepackage{color}

\usepackage{placeins}  % for FloatBarrier
\usepackage{flafter}     % Bilder immer nach \figure Befehl

\usepackage{color}

\newcommand{\gnat}{\ensuremath{\gamma_{\text{nat}}}}
\newcommand{\ginv}{\ensuremath{\gamma_{\text{inv}}}}
\newcommand{\gtyp}{\ensuremath{\gamma_{\text{typ}}}}

\makeatletter
\let\Hy@backout\@gobble
\makeatother

\begin{document}

\title{
Chaotic resonance modes in dielectric cavities:
\\
Product of conditionally invariant measure and universal fluctuations
}

\author{Roland Ketzmerick}
\affiliation{Technische Universit\"at Dresden,
    Institut f\"ur Theoretische Physik and Center for Dynamics,
    01062 Dresden, Germany}

\author{Konstantin Clau{\ss}}
\affiliation{Technische Universit\"at Dresden,
    Institut f\"ur Theoretische Physik and Center for Dynamics,
    01062 Dresden, Germany}
\affiliation{Department of Mathematics, Technical University of Munich, Boltzmannstr. 3, 85748 Garching, Germany}

\author{Felix Fritzsch}
\affiliation{Technische Universit\"at Dresden,
    Institut f\"ur Theoretische Physik and Center for Dynamics,
    01062 Dresden, Germany}
\affiliation{Physics Department, Faculty of Mathematics and Physics,
	University of Ljubljana, Ljubljana, Slovenia}

\author{Arnd B\"acker}
\affiliation{Technische Universit\"at Dresden,
    Institut f\"ur Theoretische Physik and Center for Dynamics,
    01062 Dresden, Germany}

\date{\today}
\begin{abstract}
We conjecture that chaotic resonance modes in scattering systems are a product of 
a conditionally invariant measure from classical dynamics and
universal exponentially distributed fluctuations.
The multifractal structure of the first factor depends strongly on the lifetime of the mode
and describes the average of modes with similar lifetime.
The conjecture is supported for a dielectric cavity with chaotic ray dynamics
at small wavelengths,
in particular for experimentally relevant modes
with longest lifetime.
We explain scarring of the vast majority of modes along segments of rays
based on multifractality and universal fluctuations,
which is conceptually different from periodic-orbit scarring.
\end{abstract}

\maketitle

Eigenfunctions in closed quantum systems with classically chaotic dynamics,
e.g., in quantum billiards,
are well understood based on 
quantum ergodicity, 
universal fluctuations,
and scarring
along unstable periodic orbits~\cite{HaaGnuKus2018, Sto1999}.
Resonance modes in chaotic scattering systems 
with escape of particles~\cite{AltPorTel2013, Gas2014b, Smi1989}, 
e.g., the paradigmatic three-disk scattering~\cite{GasRic1989c, Wir1999, WeiBarKuhPolSch2014},
have a fractal support on which they
are distributed depending on their 
lifetime~\cite{CasMasShe1999b, KeaNovPraSie2006, NonRub2007, KeaNonNovSie2008, ErmCarSar2009,ClaKoeBaeKet2018, BilGarGeoGir2019}
and the spectrum follows a fractal Weyl 
law~\cite{Sjo1990, Lin2002, LuSriZwo2003, SchTwo2004}.

\begin{figure}[b!]
	\vspace*{-0.4cm}
	\includegraphics[scale=1.0]{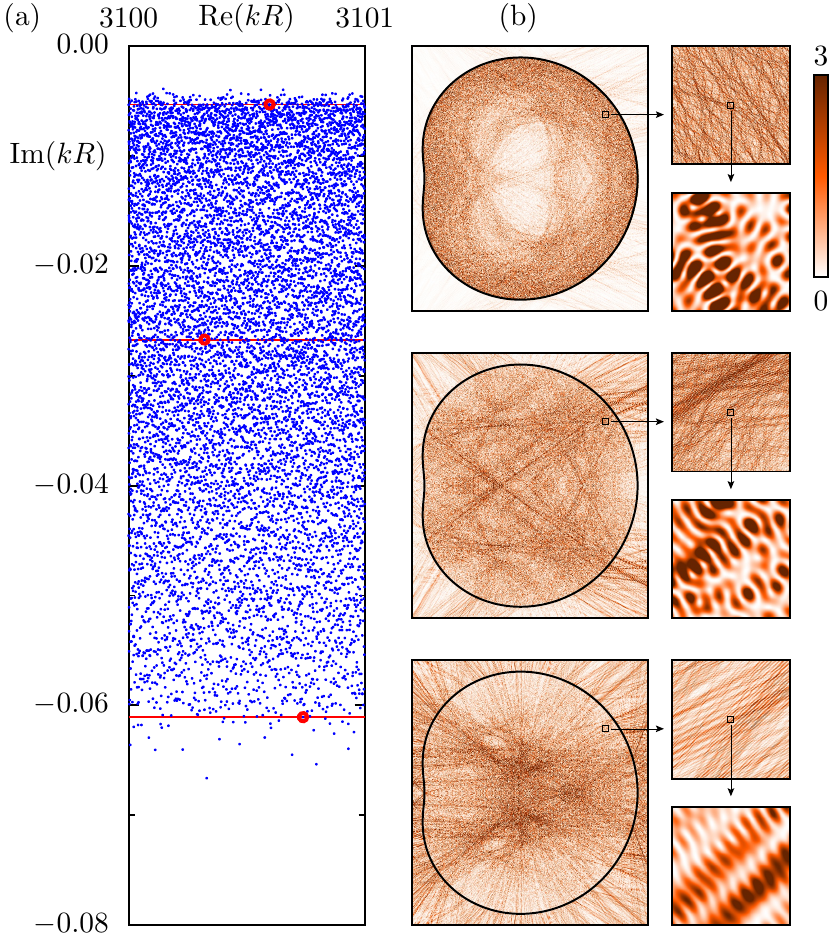}
	\caption{
		(a) Spectrum of 9964 antisymmetric TM polarized modes 
		of a lima\c{c}on cavity at small wavelengths.
		Three poles highlighted closest to classical decay rates
		$\gnat$, $\gtyp$, and $\ginv$ (lines, top to bottom).
		(b) Corresponding modes 
		showing strongly different overall intensity structures.
		Two consecutive magnifications, each by a factor 20, 
		resolve the wavelength.
	}
	\label{FIG:spectrum_individual}
\end{figure}

Resonance modes in scattering systems with partial escape of probability~\cite{AltPorTel2013}
are less understood. 
The most relevant example are dielectric 
microcavities~\cite{CaoWie2015},
see Fig.~\ref{FIG:spectrum_individual}.
The relation of cavity shape, ray dynamics, mode structure, and far-field emission
pattern has been studied extensively
experimentally and theoretically~\cite{NoeSto1997, GmaCapNarNoeStoFaiSivCho1998,
	LeeLeeChaMooKimAn2002,
	LeeRimRyuKwoChoKim2004,
    SchRexTurChaStoBenZys2004, 
    FanCaoSol2007, 
    FanCao2007,
    TanHenFukHar2007,
    ShiHar2007,
    WieHen2008,
	SonFanLiuHoSolCao2009,
	ShiHenWieSasHar2009, 
	ShiHarFukHenSasNar2010,
	XiaZouLiDonHanGon2010,
	AlbHopEbeArnEmmSchHoeForKamWieRei2012,
	HarShi2015,
	SunShiFukHar2016,  
    KulWie2016,
	BitKimZenWanCao2020}. 
A multifractal probability distribution based on ray dynamics~\cite{LeeRimRyuKwoChoKim2004},
a so-called natural conditionally invariant measure~\cite{DemYou2006},
gives a good description of modes with long 
lifetimes~\cite{LeeRimRyuKwoChoKim2004, ShiHar2007, WieHen2008,
ShiHenWieSasHar2009, ShiHarFukHenSasNar2010, HarShi2015, KulWie2016, BitKimZenWanCao2020}.

However, resonance modes have various lifetimes, while the natural measure
applies to precisely one lifetime.
It is not understood how the 
multifractal structure of resonance modes
depends on their lifetime,
even in the simplest setting of a cavity shape with fully chaotic ray dynamics.
Experimentally this is most relevant for modes with
the longest lifetimes,
which have longer lifetimes than the natural measure.
Their enhanced scarring has been reported experimentally and 
numerically~\cite{LeeLeeChaMooKimAn2002, GmaNarCapBaiCho2002, HarFukDavVacMiyNisAid2003, FanCaoSol2007, FanCao2007, WisCar2008, NovPedWisCarKea2009, Nov2013, CarBenBor2016, PraCarBenBor2018},
but the relation to periodic-orbit scarring is under investigation.
On a fundamental level it is open, 
which features of a resonance mode are
system specific with a ray-dynamical origin
and which are universal wave phenomena.

In this Letter we answer these questions based on recent progress 
on quantum maps with partial escape~\cite{ClaAltBaeKet2019, ClaKunBaeKet2021}, 
leading to the following conjecture:
\textit{Chaotic resonance modes in scattering systems 
	are a product}
	\begin{equation}\label{eq:conjecture}
		|\psi(\boldsymbol{r})|^2 = \varrho(\boldsymbol{r}) \cdot  \eta(\boldsymbol{r})
	\end{equation}	
\textit{of
	(i) conditionally invariant measures from classical dynamics 
	with a smoothed spatial density $\varrho(\boldsymbol{r})$
	depending on the mode's lifetime
	and
	(ii) universal 
	exponentially distributed fluctuations $\eta(\boldsymbol{r})$ with mean one.}
An immediate consequence is that the average intensity of modes 
with similar lifetime is determined by the first factor.
We support this conjecture using a dielectric cavity with chaotic ray dynamics,
by factorizing modes into an average of modes with similar lifetime and universal fluctuations.
It is demonstrated that the multifractal structure of the average strongly depends on the lifetime. 
This is described by 
appropriate conditionally invariant measures based on ray dynamics.
In particular, this holds for the experimentally relevant 
modes of optical microcavities with longest lifetime.
We explain the scarring of the vast majority of modes along segments of rays 
based on multifractality and universal fluctuations.
It conceptually differs from periodic-orbit scarring and
becomes even more prominent in the semiclassical limit.
Our computations are done at very small wavelengths.

%%%%%%%%%%%%%%%%%%%%%%%%%%%%%%%%%%%%%%%%%%%%%%%%%%%%%%%%%%%%%%%%%%%%%%%%%%%%%
\vspace*{0.1cm}
\emph{Modes of dielectric cavity.}---%
%%%%%%%%%%%%%%%%%%%%%%%%%%%%%%%%%%%%%%%%%%%%%%%%%%%%%%%%%%%%%%%%%%%%%%%%%%%%%
We study passive modes in a lima\c{c}on shaped cavity, 
given in polar coordinates by
$\rho(\varphi) = R (1 + \varepsilon\cos \varphi)$~\cite{Rob1983,WieHen2008}.
For $\varepsilon = 0.6$ it is non-convex
and shows chaotic ray dynamics practically
everywhere in phase space, with
possible regular regions~\cite{DulBae2001} being negligibly small.
We choose a refractive index $n=3.3$ typical for a semiconductor laser cavity~\cite{CaoWie2015}
and outside the cavity $n=1$.
Antisymmetric TM polarized modes
$\psi(\boldsymbol{r})$
fulfilling the Helmholtz equation
$[\Delta + n^2(\boldsymbol{r}) k^2] \psi(\boldsymbol{r}) = 0$
with outgoing boundary conditions
are computed for complex wave numbers with
$\text{Re}(kR) \in [3100, 3101]$ and $\text{Im}(kR) \in [-0.1, 0]$.
With $nkR \approx 10^{4}$
this is more than an order of magnitude further in the short wavelength limit
than previous studies of dielectric cavities, see e.g., 
Refs.~\cite{CaoWie2015, SunShiFukHar2016, KulWie2016, BitKimZenWanCao2020}.
This allows for numerical comparison to wave chaos experiments
with large cavities~\cite{BitKimZenWanCao2020}.

One observes a band of resonance poles  
with two spectral gaps, one near the real line 
and one further in the complex plane,
see Fig.~\ref{FIG:spectrum_individual}(a).
The upper end of the spectrum, with long-lived modes of high quality factor 
$Q = -(1/2) \text{Re}(k) / \text{Im}(k) \approx 4 \cdot 10^5$,
occurs near the
classical natural decay rate $\gnat$
from ray dynamics, discussed below.
The lower end of the spectrum, with short-lived modes 
and a gap towards external modes~\cite{DetMorSieWaa2009},
occurs near the classical natural decay rate $\ginv$ from the inverted ray dynamics,
see below.
The middle of the spectrum corresponds to the typical classical decay rate $\gtyp$
of an ergodic ray~\cite{NonSch2008}. 
These classical decay rates are indicated in Fig.~\ref{FIG:spectrum_individual}(a)
by horizontal lines at
$\text{Im}(kR) = - (\gamma / 2) R / c$.

\begin{figure*}[t!]
	\includegraphics[scale=1.0]{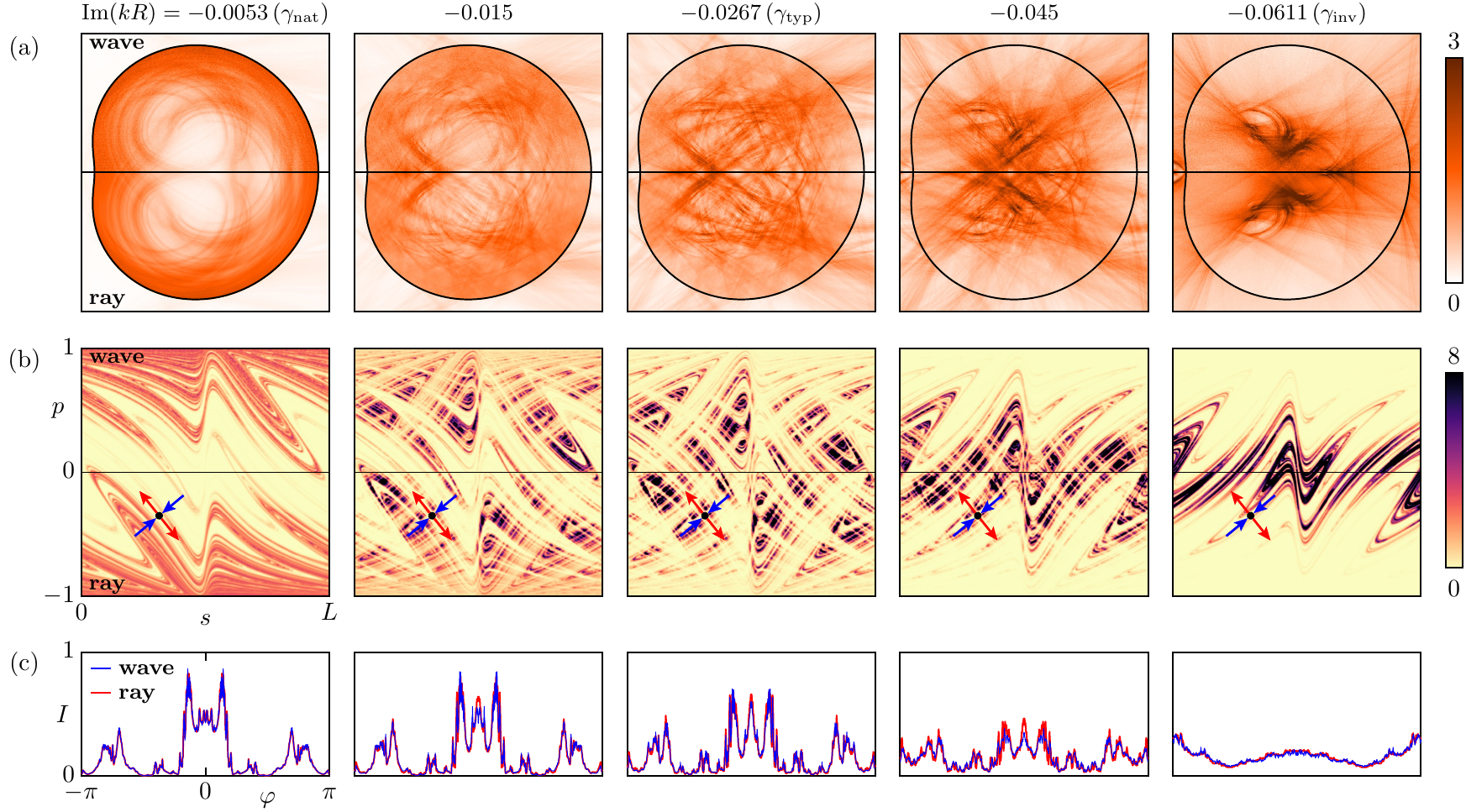}
	\caption{
		Average of modes compared to ray dynamics.
		(a) Upper half (wave): Average intensity of 200 modes 
		each nearest to indicated $\text{Im}(kR)$,
		with first and last corresponding
		to  decay rates $\gnat$ and $\ginv$. 
		Lower half (ray):
		Corresponding 
		smoothed spatial density $\varrho(\boldsymbol{r})$ from the
		conditionally invariant product measure for $\xi = 0, 0.232, 0.432, 0.722, 1$
		(left to right).
		(b) Same as (a) for average incident Husimi function.
		Unstable and stable direction shown for exemplary point in phase space.
		(c) Average normalized far-field intensity (thin, blue) compared to ray calculation (thick, red).
	}
	\label{FIG:overview_wave_ray}
\end{figure*}

For exemplary modes with long, medium, and short lifetimes 
the intensity $|\psi(\boldsymbol{r})|^2$ 
is presented on a $500 \times 560$ grid in Fig.~\ref{FIG:spectrum_individual}(b),
showing large-scale structures and fluctuations.
Consecutive magnifications reveal finer structures and
eventually a smooth wave function on the scale of the wavelength.
As expected, one observes a drastic change of the overall structure
from long-lived to short-lived modes,
with strong intensities shifting from near the boundary 
to the center of the cavity
and angles of reflection changing from total internal reflection to 
perpendicular to the boundary (see also Fig.~\ref{FIG:overview_wave_ray}(b) below).
For modes with a similar lifetime one finds similar structures,
see gallery of modes~\cite{SM}.

\begin{figure}[b!]
	%\vspace*{-0.4cm}
	\includegraphics[scale=1.0]{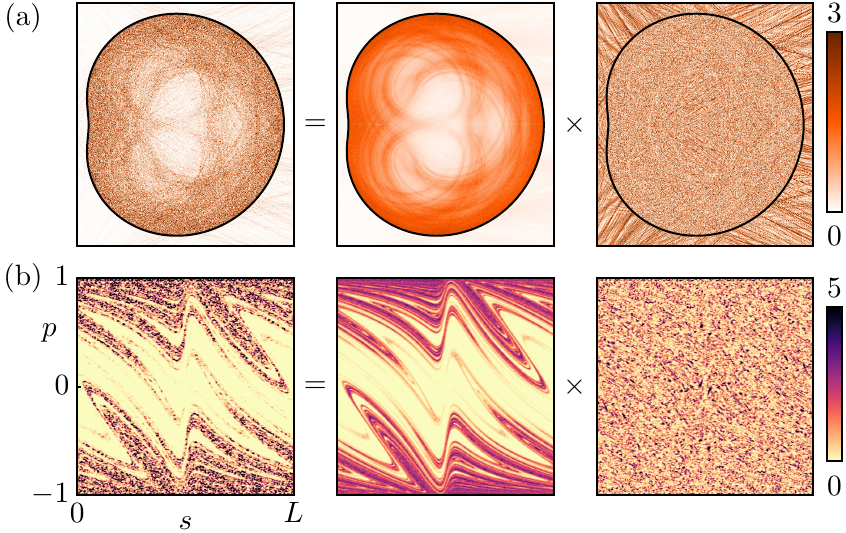}
	\caption{(a) Factorization of intensity 
		$|\psi(\boldsymbol{r})|^2$ (left)
		in position space for long-lived mode 
		from Fig.~\ref{FIG:spectrum_individual}(b) 
		into average 
		$\langle |\psi(\boldsymbol{r})|^2 \rangle$ (middle)
		of 200 modes nearest to
		$\text{Im}(kR) = -0.0053$ ($\gnat$)
		and fluctuations
		$\eta(\boldsymbol{r})$ (right). 
		(b) Factorization of incident Husimi function 
		$H(s,p)$
		for same mode.
		In all figures
		the average intensity (in position space within the cavity) is scaled to one and
		intensities greater than the maximal value of the color bar 
		are shown with darkest color.
	}
	\label{FIG:factorization}
\end{figure}

Numerically, we determine the modes 
using boundary integral equations~\cite{KniRei1996, Wie2003}.
For analytic boundaries we use the approach of Ref.~\cite{Kre1991},
which allows to use
just slightly more than two discretization points per wavelength on the boundary,
giving modes and spectrum with machine accuracy.
Here we need $N_b = 11500$ points 
on the desymmetrized boundary of length $L/2$ with $L = 6.8627 R$
at wavelength $\lambda = 2\pi /\text{Re}(nk)$ giving
$b = 2 N_b \lambda / L = 2.06 $ points per wavelength.
We find all poles in a complex wave number region
using a Taylor expansion of the matrix equation in $k$,
extending an approach for quantum billiards~\cite{VebProRob2007, PeiDieHua2019}
to complex $k$.
We increase the accuracy to machine precision by
applying the method of Ref.~\cite{VebProRob2007} to every pole in a subspace
and subsequent convergence steps.
Further details about the numerical approach will be published elsewhere~\cite{Ket2022:p}.
The high precision of this approach has been used for analyzing resonance assisted tunneling
with a resolution of $\text{Im}(kR) = 10^{-12}$~\cite{FriKetBae2019}.
We expect to find all poles in the considered complex wave number region,
namely 9964,
which is supported by the leading order Weyl term
of a dielectric cavity~\cite{BogDubSch2008, BogDjeDubGozLebSchUlyZys2011} giving 
\mbox{$N = \frac{A}{4\pi} n^2 (3101^2-3100^2) = 9960.5$} 
poles, with area $A = (\pi/2) (1 + \varepsilon^2/2)$ of the desymmetrized cavity. 
The next order boundary term for TM modes~\cite{BogDubSch2008, BogDjeDubGozLebSchUlyZys2011}
gives a contribution of less than one mode.

%%%%%%%%%%%%%%%%%%%%%%%%%%%%%%%%%%%%%%%%%%%%%%%%%%%%%%%%%%%%%%%%%%%%%%%%%%%%%
\vspace*{0.1cm}
\emph{Factorization.}---%
%%%%%%%%%%%%%%%%%%%%%%%%%%%%%%%%%%%%%%%%%%%%%%%%%%%%%%%%%%%%%%%%%%%%%%%%%%%%%
We numerically extract the two factors of Eq.~(\ref{eq:conjecture})
from the resonance modes.
The average intensity 
$\varrho(\boldsymbol{r}) = \langle |\psi(\boldsymbol{r})|^2 \rangle$
is determined from the 200 modes nearest in lifetime. The fluctuations at every point of the grid
are determined by 
$\eta(\boldsymbol{r}) = |\psi(\boldsymbol{r})|^2 / \langle |\psi(\boldsymbol{r})|^2 \rangle$.
For the long-lived mode of Fig.~\ref{FIG:spectrum_individual}(b)
this factorization  
in position space is visualized in Fig.~\ref{FIG:factorization}(a).

Such a factorization also applies to
the incident Husimi function $H(s,p)$~\cite{HenSchSch2003}
on the boundary phase space $(s, p)$,
where $s=0$ is the boundary point for $\varphi = 0$
and $p$ is the normalized momentum parallel to the boundary.
The average Husimi function 
$\varrho(s,p) = \langle H(s,p) \rangle$
is determined from the 200 modes nearest in lifetime. The fluctuations at every point $(s,p)$
are determined by 
$\eta(s,p) = H(s,p) / \langle H(s,p) \rangle$.
This factorization is visualized in Fig.~\ref{FIG:factorization}(b).

One observes
that the fluctuations $\eta$ in position space 
are quite uniform and have almost no spatial structure
within the cavity.
The same is true for the fluctuations $\eta$
in the boundary phase space.
This even holds for regions, where the intensity of the mode 
and the average are both close to zero.
More generally, we expect that the same factorization holds 
in the full phase space restricted to the energy shell.
In the following we will
demonstrate that the two factors fulfill
the conjecture.

\begin{figure}[t!]
	\includegraphics[scale=1.0]{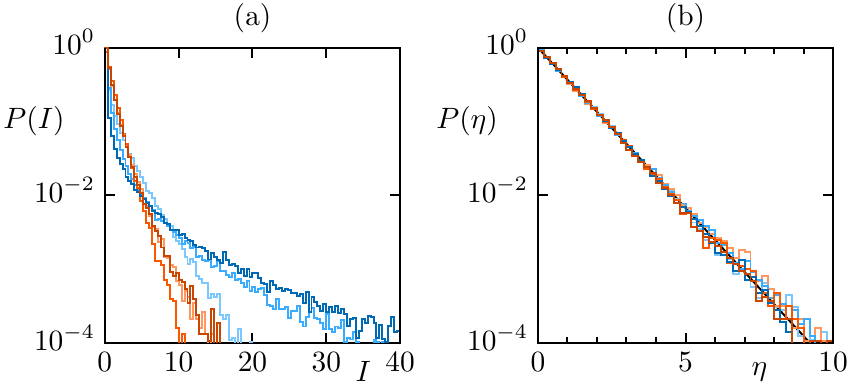}
	\caption{
		(a)
		Distribution of intensities $I$ in position space inside cavity (red) 
		and boundary phase space (blue)
		for the three modes of Fig.~\ref{FIG:spectrum_individual}(b) 
		(light to dark for decreasing lifetime) each with mean intensity one.
		(b)
		Same as (a) for fluctuations~$\eta$ compared to a universal exponential distribution 
		$P(\eta)=\exp(-\eta)$ (black).
	}
	\label{FIG:histogram}
\end{figure}

%%%%%%%%%%%%%%%%%%%%%%%%%%%%%%%%%%%%%%%%%%%%%%%%%%%%%%%%%%%%%%%%%%%%%%%%%%%%%
\vspace*{0.2cm}
\emph{Universal fluctuations.}---%
%%%%%%%%%%%%%%%%%%%%%%%%%%%%%%%%%%%%%%%%%%%%%%%%%%%%%%%%%%%%%%%%%%%%%%%%%%%%%
In Fig.~\ref{FIG:histogram}(a)
the non-universal distribution of the intensities $I$
in position space, $I=|\psi(\boldsymbol{r})|^2$,
and phase space, $I=H(s,p)$, 
is shown for the three modes from
Fig.~\ref{FIG:spectrum_individual}(b).
In contrast, in Fig.~\ref{FIG:histogram}(b)
the fluctuations $\eta$
follow a universal exponential distribution of mean one
for more than three orders of magnitude in all cases.
This has been conjectured for
quantum maps with partial escape~\cite{ClaKunBaeKet2021},
corresponds to the properties of a normalized complex random
vector~\cite{BroFloFreMelPanWon1981, KusMosHaa1988},
and supports 
property (ii) of the conjecture.
Accordingly, the complex 
amplitude fluctuations $\psi(\boldsymbol{r}) / \sqrt{\langle |\psi(\boldsymbol{r})|^2 \rangle}$
follow a complex Gaussian distribution of variance one (not shown).  
We stress, that the analysis of the fluctuations presented here
is possible only, if the 
average is determined from
sufficiently many and sufficiently nearby modes in lifetime.
We mention that the fluctuations $\eta$
of the far-field intensities 
also agree with the universal exponential distribution (not shown).
The correlations of $\eta(\boldsymbol{r})$ and $\eta(s,p)$
on the scale of the wavelength and under time evolution are expected to
show similar behavior as for closed systems~\cite{Sch2005b}.

A factorization of chaotic modes into an average part and a fluctuating part
is used for the explanation of single-mode lasing~\cite{HarSunShi2017},
however under the strong assumption of small fluctuations. 
For the analysis of
single- versus multi-mode lasing the spatial overlap of two modes within the cavity, 
$C = \int |\psi_1(\boldsymbol{r})| \, |\psi_2(\boldsymbol{r})| \, \text{d} \boldsymbol{r}$,
is important~\cite{SunShiFukHar2016},
with $\int |\psi_{1,2}(\boldsymbol{r})|^2 \, \text{d} \boldsymbol{r} = 1$.
For long-lived modes of the stadium billiard the
mean value is $C \approx 0.77$~\cite{SunShiFukHar2016}.
From the above universal fluctuations and assuming independence from the average
we find universally $C=\pi/4 \approx 0.785$ for two chaotic modes with 
nearby lifetimes
in any chaotic cavity. This is consistent with the findings in Ref.~\cite{SunShiFukHar2016}
and is numerically well confirmed in the present cavity.
Thus the demonstrated factorization has a strong impact
on the experimentally relevant question of single-mode lasing.

%%%%%%%%%%%%%%%%%%%%%%%%%%%%%%%%%%%%%%%%%%%%%%%%%%%%%%%%%%%%%%%%%%%%%%%%%%%%%
\vspace{0.3cm}
\emph{Average of modes and ray dynamics.}---%
%%%%%%%%%%%%%%%%%%%%%%%%%%%%%%%%%%%%%%%%%%%%%%%%%%%%%%%%%%%%%%%%%%%%%%%%%%%%%
In the following we demonstrate that the average structure of chaotic modes
with similar lifetime
strongly depends on the lifetime and 
is described by
appropriate conditionally invariant measures of ray dynamics,
supporting property (i) of the conjecture.
So far averages have been computed for long-lived 
modes~\cite{ShiHenWieSasHar2009, HarShi2015, KulWie2016, BitKimZenWanCao2020}.
In Fig.~\ref{FIG:overview_wave_ray} we show the strong dependence on $\text{Im}(kR)$ 
for the average intensity in position space, phase space, and in the far field.
The averages are each over 200 nearby modes in $\text{Im}(kR)$
normalized within the cavity.
They show very fine details compared to
the individual modes
in Fig.~\ref{FIG:spectrum_individual}(b).

In Fig.~\ref{FIG:overview_wave_ray}(b) we show the corresponding incident Husimi functions.
Their structure changes completely with $\text{Im}(kR)$.
One observes
fractal structures in both the stable and the unstable direction of the ray dynamics.
The average far-field intensity 
is presented in Fig.~\ref{FIG:overview_wave_ray}(c).
It shows strong directionality in agreement with Ref.~\cite{WieHen2008}.

These averaged modes are well explained by conditionally invariant measures
based on ray dynamics
and smoothed on the scale of a wave length,
giving spatial densities $\varrho(\boldsymbol{r})$ 
and densities $\varrho(s,p)$ in the boundary phase space.
The wave-ray comparison supports
property (i) of the conjecture,
see upper half (wave) and lower half (ray) in Fig.~\ref{FIG:overview_wave_ray}(a) and (b). 
We find perfect agreement at 
the natural decay rate $\gnat$ and
the inverse natural decay rate $\ginv$,
while for 
all other decay rates we use the approximate,
but very good,
 description
by product measures,
as described in the following.

The natural conditionally invariant measure~\cite{DemYou2006} 
with natural decay rate $\gnat$ is determined from
time evolution of a smooth initial density in phase space using ray dynamics 
and intensity changes at each reflection according to Fresnel's laws.
This approach has been established for microcavities by 
Soo-Young Lee et al.~\cite{LeeRimRyuKwoChoKim2004}
and is confirmed for many chaotic 
cavities~\cite{LeeRimRyuKwoChoKim2004, ShiHar2007, WieHen2008,
	ShiHenWieSasHar2009, ShiHarFukHenSasNar2010, HarShi2015, KulWie2016, BitKimZenWanCao2020}.
We stress that this measure describes those long-lived modes only,
which are close to the natural decay rate $\gnat$.
Note that at $\gnat$ the phase-space distribution is
smooth along the unstable direction, 
see Fig.~\ref{FIG:overview_wave_ray}(b, left).

A second natural measure is determined from the inverse dynamics,
i.e.\ applying the inverse of Fresnel's laws at each reflection~\cite{AltPorTel2015,GutOsi2015,ClaAltBaeKet2019}.
The corresponding natural decay rate of the inverse dynamics, $\ginv$,
corresponds to short-lived modes, which again are perfectly described,
see Fig.~\ref{FIG:overview_wave_ray} right.
Note that at $\ginv$ the phase-space distribution is 
smooth along the stable direction.

\begin{figure}[b!]
	\includegraphics[scale=1.0]{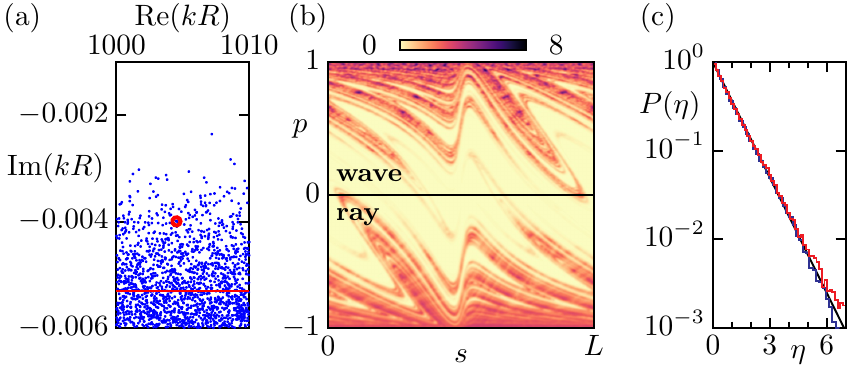}
	\caption{
		(a) Upper part of spectrum including 898 long-lived modes 
		with $\text{Im}(kR) > -0.0053$ (horizontal line, $\gnat$)
		for $\text{Re}(kR) \in [1000, 1010]$.
		(b) Upper half (wave):  
		Average incident Husimi function for 100 modes nearest to $\text{Im}(kR) = -0.004$.
		Lower half (ray):
		Corresponding 
		smoothed density $\varrho(s,p)$ based on the product measure
		for $\xi = -0.06$.
		(c) As Fig.~\ref{FIG:histogram}(b) for mode nearest to $\text{Im}(kR) = -0.004$ marked in~(a).
	}
	\label{FIG:supersharp}
\end{figure}

For all other decay rates we use the 
product measures introduced for quantum maps~\cite{ClaAltBaeKet2019}.
They are based on the observation that locally in phase space
the averaged modes have an (approximate) product structure along stable and unstable direction
of ray dynamics.
The product measures interpolate between the natural and the inverse natural measure
(depending on a parameter $\xi \in \mathbb{R}$)
and provide conditionally invariant measures for all decay rates,
see Ref.~\cite{ClaAltBaeKet2019} for their construction.
These product measures
show very good, but not perfect, agreement with the averaged modes, 
see the three intermediate examples in Fig.~\ref{FIG:overview_wave_ray}.
Thus we have found wave-ray correspondence for the multifractal
structures of the average of chaotic modes down to unprecedented fine details.

%%%%%%%%%%%%%%%%%%%%%%%%%%%%%%%%%%%%%%%%%%%%%%%%%%%%%%%%%%%%%%%%%%%%%%%%%%%%%
\vspace{0.3cm}
\emph{Modes with longest lifetime.}---%
%%%%%%%%%%%%%%%%%%%%%%%%%%%%%%%%%%%%%%%%%%%%%%%%%%%%%%%%%%%%%%%%%%%%%%%%%%%%%
Experimentally, the most relevant modes for lasing
are those with 
longest lifetime, i.e.\ closest to the real axis.
Their spectral density decreases with smaller wavelength~\cite{Nov2012, GutOsi2015}.
In order to have enough modes for averaging,
it is therefore numerically convenient to use larger wavelengths, 
see spectrum in Fig.~\ref{FIG:supersharp}(a).
The average incident Husimi function 
near $\text{Im}(kR) = -0.004$
has structure along the unstable direction, 
see Fig.~\ref{FIG:supersharp}(b, upper half),
and thus
clearly differs from the one at $\gnat$,
which is smooth along the unstable direction,
see Fig.~\ref{FIG:overview_wave_ray}(b, left).
This structure is qualitatively well described by the
corresponding conditionally invariant product measure,
see Fig.~\ref{FIG:supersharp}(b, lower half).
We analyze the relative fluctuations and find
a universal exponential distribution over almost three orders of magnitude, 
see Fig.~\ref{FIG:supersharp}(c).
Thus factorization into ray-dynamical average and universal fluctuations
is essential for understanding the structure of these modes.

As an aside we mention that for the considered cavity shape
there are whispering gallery-like modes 
for $\text{Re}(kR) \approx 1000$ and below,
which are related to partial barriers.
For larger $\text{Re}(kR) \approx 3000$ this ray-dynamical property does no longer
affect the modes, as expected from universal scaling 
properties~\cite{MicBaeKetStoTom2012, KoeBaeKet2015}.

%%%%%%%%%%%%%%%%%%%%%%%%%%%%%%%%%%%%%%%%%%%%%%%%%%%%%%%%%%%%%%%%%%%%%%%%%%%%%
\vspace{0.1cm}
\emph{Scarring.}---%
%%%%%%%%%%%%%%%%%%%%%%%%%%%%%%%%%%%%%%%%%%%%%%%%%%%%%%%%%%%%%%%%%%%%%%%%%%%%%
The scarring of eigenfunctions in closed chaotic quantum billiards
refers to an enhancement along short unstable periodic orbits~\cite{Hel1984, Kap1999}.
For chaotic modes in dielectric cavities and in corresponding
quantum maps with escape enhanced scarring of modes has been 
reported~\cite{LeeLeeChaMooKimAn2002, GmaNarCapBaiCho2002, HarFukDavVacMiyNisAid2003, FanCaoSol2007, FanCao2007, WisCar2008, NovPedWisCarKea2009, Nov2013, CarBenBor2016, PraCarBenBor2018}.
We observe at very small wavelengths that, in fact, 
the vast majority of modes show enhanced intensities along segments of rays.
This is visible in Fig.~\ref{FIG:spectrum_individual}(b) 
for examples with medium and short lifetime 
as well as in the gallery of modes~\cite{SM}
for modes with longest lifetime.
It is best seen when the mode is shown with a resolution 
on the scale of the wavelength,
see the first magnifications in Fig.~\ref{FIG:spectrum_individual}(b)
and also see Fig.~S7~\cite{SM}, which shows
the mode in the middle of Fig.~\ref{FIG:spectrum_individual}(b)
with a tenfold finer resolution.

We explain this type of scarring in scattering systems
based on multifractality and universal fluctuations
and emphasize that it conceptually differs from periodic-orbit scarring.
It has a combined ray and wave origin: 
Whenever the multifractal average structure (ray origin) 
shows strong intensity enhancements in phase space,
then the additional universal fluctuations (wave origin) 
give rise to some phase-space points with extremely high intensities. 
For the examples in Fig.~\ref{FIG:histogram}(a) there are even 
intensities that are more than a factor of 100 larger than the mean intensity.
In position space this gives rise to enhancement of the mode along 
the corresponding ray in forward and backward direction, 
sometimes persisting for one or two reflections.
Thus we call this phenomenon \textit{ray-segment scarring}. 
The most likely directions are determined
by the high intensities of the multifractal averaged structure in phase space.
The specific direction of the ray segment varies from mode to mode,
as the phase-space points with extreme intensities vary
due to the universal fluctuations,
see e.g.\ Fig.~S3~\cite{SM}.

The strongest intensity variation in the averaged modes
occurs according to
Fig.~\ref{FIG:overview_wave_ray}(b) for medium and short lifetime
and according to
Fig.~\ref{FIG:supersharp}(b) for modes with longest lifetime.
Correspondingly, the most prominent 
scarring occurs in these cases,
see gallery of modes~\cite{SM}.

For increasingly smaller wavelengths the averaged modes show finer 
multifractal structures with increasing intensity maxima.
Thus we expect that ray-segment scarring 
becomes even more prominent and is visible for longer
segments of a ray in the semiclassical limit.

%%%%%%%%%%%%%%%%%%%%%%%%%%%%%%%%%%%%%%%%%%%%%%%%%%%%%%%%%%%%%%%%%%%%%%%%%%%%%
\vspace*{0.1cm}
\emph{Outlook.}---%
%%%%%%%%%%%%%%%%%%%%%%%%%%%%%%%%%%%%%%%%%%%%%%%%%%%%%%%%%%%%%%%%%%%%%%%%%%%%%
A semiclassical theory that derives the perfect conditionally invariant measures
for modes of all lifetimes remains a future challenge.
A first step in this direction is based on a local random vector model
applicable to the randomized baker map with partial escape~\cite{ClaKet2022}.
Further support for the conjecture is expected in
chaotic scattering systems with full escape,
like the three-disk system.

%%%%%%%%%%%%%%%%%%%%%%%%%%%%%%%%%%%%%%%%%%%%%%%%%%%%%%%%%%%%%%%%%%%%%%%%%%%%%
\vspace*{0.1cm}
%\acknowledgments

We thank 
S.~Bittner, E.-M.~Graefe, T.~Harayama, M.~Hentschel, J.~Kullig, J.~L\"otfering,
M.~Lebental, T.~Prosen, J.R.~Schmidt, M.~Sieber, and J.~Wiersig 
for valuable discussions
as well as the organizers of the WOMA conference series.
Funded by the Deutsche Forschungsgemeinschaft (DFG, German Research Foundation) – 262765445.

%%%%%%%%%%%%%%%%%%%%%%%%%%%%%%%%%%%%%%%%%%%%%%%%%%%%%%%%%%%%%%%%%%%%%%%%%%%%%


\end{document}